\documentclass[conference]{IEEEtran}
\usepackage{graphicx} \usepackage{tabularx}
\usepackage{float} \usepackage{amsmath}
\usepackage{booktabs}
\usepackage{longtable}
\usepackage{multirow}
\usepackage{algorithm}
\usepackage{algpseudocode}
\usepackage{cancel}
\usepackage{subcaption}
\usepackage{multicol}
\usepackage{url}
\usepackage{balance}
\usepackage{orcidlink}
\usepackage{amssymb}
\usepackage{hyperref}
\usepackage{placeins}
\usepackage{dblfloatfix}
\hypersetup{
    colorlinks=true,
    linkcolor=black,
    filecolor=magenta,
    citecolor=black,
    urlcolor=blue,
    }

\usepackage{colortbl}
\usepackage[most]{tcolorbox}
\usepackage{xcolor}

\usepackage{listings}
\usepackage{tikz}
\usetikzlibrary{calc}
\usepackage{array}
\usepackage[most]{tcolorbox}
\usepackage{multirow}

\title{LLM-Based Discovery of Latent Requirements from Stakeholder Conversations: Preliminary Results from Industry}

\author{
\IEEEauthorblockN{
Mithila Sivakumar\IEEEauthorrefmark{1}\IEEEauthorrefmark{2},
Martin Lochner\IEEEauthorrefmark{2},
Shiva Nejati\IEEEauthorrefmark{1},
Mehrdad Sabetzadeh\IEEEauthorrefmark{1}
}
\IEEEauthorblockA{\IEEEauthorrefmark{1}EECS, University of Ottawa, Canada\\
\{msiva025, snejati, m.sabetzadeh\}@uottawa.ca
}
\IEEEauthorblockA{\IEEEauthorrefmark{2}eSentire, Canada\\
mlochner@uwaterloo.ca
}
}

\newcommand{\framework}{LENS}

\begin{document} 

\maketitle

\thispagestyle{plain}

\begin{abstract}
Stakeholder interviews are an important source of information for requirements elicitation, yet many relevant requirements remain implicit in such conversations. Stakeholders frequently describe workflows, challenges, and operational practices without explicitly articulating the software capabilities that could address them. Recent work has considered the use of LLMs to analyze conversational data and extract requirements from stakeholder interviews. Existing  approaches, however, primarily focus on identifying explicitly stated requirements, leaving implicit opportunities largely unexplored. In this paper, we present \framework\ (LLM-Enabled Needs Discovery from Stakeholder Interviews), an approach that analyzes stakeholder interview transcripts to both extract explicit requirements and infer additional latent requirements. \framework\ performs this inference by reasoning over stakeholder statements together with contextual information about organizational tools and infrastructure. Both extracted and inferred requirements are represented as user stories and linked to transcript excerpts to ensure traceability. We conduct a preliminary evaluation of \framework\ using twelve stakeholder interview transcripts collected in an industrial setting involving cybersecurity operations. We show that \framework\ achieves an average F1-score of 84.4\% for extracting explicit requirements, while, on average, 75\% of the latent requirements identified by \framework\ were perceived as providing useful automation or time-saving potential by domain experts.
\end{abstract}

\begin{IEEEkeywords}
Conversational RE, Requirements Elicitation, Large Language Models (LLMs),  Latent Requirements.
\end{IEEEkeywords} \section{Introduction}\label{sec:intro}
Stakeholder conversations are a key source of requirements elicitation data~\cite{reconsum2023, voria2025recover}. Recent research  suggests that Large Language Models (LLMs) can help transform such conversational data, e.g., interview transcripts, into software requirements~\cite{voria2025recover, almeida2025elicitation}. However, current approaches focus primarily on identifying and consolidating \emph{explicitly stated} requirements. This is limiting in practice: stakeholders do not always articulate their needs directly, and important requirements may only become visible when stakeholders' statements are interpreted in light of existing workflows, organizational tools, and available automation infrastructure~\cite{burnay2014stakeholders}.

In this paper, we introduce \textit{\framework} (\textbf{L}LM-\textbf{E}nabled \textbf{N}eeds Discovery from \textbf{S}takeholder Interviews), an approach that infers \textbf{latent requirements} from interview transcripts. We define latent requirements as \textbf{plausible opportunities for introducing software that stakeholders do not explicitly articulate, but that become apparent when their statements are interpreted within the context of existing organizational tools, workflows, and infrastructure.} In addition, \framework\ extracts requirements that are expressly stated in the conversation and presents both the extracted and inferred requirements as structured \emph{user stories}~\cite{qus}. Each user story is linked to the corresponding transcript excerpt and timestamp to ensure traceability to the original stakeholder statements.

Our work is based on the hypothesis that LLMs can support requirements elicitation not only by extracting explicitly stated requirements, but also by reasoning over stakeholder descriptions of work practices, inefficiencies, and pain points to identify candidate requirements that stakeholders do not articulate. Guided by this hypothesis, \framework\ operates in two steps. First, it extracts requirements that are expressly stated in interview transcripts. Second, it analyzes those transcripts together with information about the organization's operational context to infer latent requirements. 

For each inferred requirement, \framework\ provides a structured rationale that explains the underlying problem, how the requirement addresses it within the organization, and the expected impact on organizational practices and outcomes.

\begin{figure*}[!t]
\centering
\fbox{
\begin{minipage}{0.95\linewidth}
\ttfamily
\small

...... \newline

\textcolor{red}{[Interviewer]}:6:48 Let’s map out exactly how an analyst performance assessment and reporting flows through your team, who does what, when, and how decisions get made. \newline

.....\newline

\textcolor{red}{[Interviewee (role redacted)]}: 19:26 \textcolor[HTML]{3333FF}{...it would be really helpful to bring together analyst performance insights in one place. Ideally, it would highlight strengths, areas for improvement, and trends over time. For instance, identifying recurring question patterns from \textcolor{red}{[Internal Messaging Platform]} would be a great starting point. That kind of automation would save time and allow team leads to focus more on supporting and developing their team.} 
\newline
\newline
......
\end{minipage}
}
\caption{Example stakeholder interview transcript from eSentire. Redacted text appears in square brackets ([]) and is highlighted in \textcolor{red}{red}. Text highlighted in \textcolor{blue}{blue} marks explicit automation intent expressed by the interviewee.}
\label{fig:transcript}
\end{figure*}

We conduct a preliminary evaluation of \framework\ through an industrial collaboration with \emph{eSentire}, focusing on workflow automation requirements in a Security Operations Centre (SOC), where analysts monitor, investigate, and respond to cybersecurity threats. Our study uses a dataset of twelve stakeholder interview transcripts and evaluations from two domain experts at eSentire. We address two research questions (RQs). RQ1 examines the \textit{accuracy} of extracting expressly stated requirements from stakeholder interview transcripts. RQ2 evaluates the \textit{usefulness} of inferred latent requirements in terms of perceived time savings and practical value of it. Our results show that, for explicitly stated requirements, \framework\ yields a precision of 73\%, a recall of 100\%, and an F1-score of 84.4\%. For latent requirements, on average, 75\% were rated as providing useful automation or time-saving potential by the domain experts, indicating their perceived usefulness as early-stage elicitation artifacts rather than \hbox{implementation-ready requirements.}

\textbf{Contributions.}
Our main contributions are as follows: (1)~\framework, an LLM-based approach that extracts explicitly stated requirements and infers latent ones from stakeholder interview transcripts by reasoning over organizational  context; and (2)~a preliminary evaluation with eSentire using twelve stakeholder interview transcripts and practitioner assessments of extraction accuracy and inferred-requirement usefulness.

The remainder of this paper is organized as follows. Section~\ref{sec:context} describes the context and motivation. Section~\ref{sec:approach} presents our approach. Section~\ref{sec:empirical_eval} details the technical aspects of the approach, reports the empirical evaluation results and discusses threats to validity. Section~\ref{sec:relwork} compares with related work. Section~\ref{sec:lessons} outlines the lessons learned. Section~\ref{sec:conclusion} concludes the paper. Section~\ref{sec:appendix} (Appendix) presents our prompts and the user-story quality criteria embedded within them. 
 \section{Industry Context \& Motivating Example}
\label{sec:context}

eSentire, Inc. (\url{https://www.esentire.com}) is a global cybersecurity company that provides managed detection and response services. The company's mandate is to use a combination of AI-driven techniques and human expertise to detect, investigate, and stop threats before they disrupt business operations. 
To support this goal, requirements are gathered through stakeholder interviews as part of a software development lifecycle built on agile principles such as iterative development and continuous stakeholder feedback.

Analyzing lengthy stakeholder interview transcripts and identifying relevant requirements remains a time-consuming manual task for requirements engineers. In addition, focusing only on explicitly stated requirements may overlook latent opportunities that emerge indirectly through discussions of repetitive activities, and inefficiencies in existing processes. Motivated by these challenges, and in collaboration with eSentire, we developed \framework\ to support analysts by automatically extracting explicitly stated requirements while also discovering plausible latent requirements that can stimulate further stakeholder discussion and iterative requirements elicitation. The following example illustrates how \framework\ processes stakeholder conversations to identify both explicit and latent requirements from real interview data.

Figure~\ref{fig:transcript} shows an excerpt of an interview transcript  
where the participants discuss about their current performance assessment and coaching process. In this interview transcript, and in other examples throughout the paper, text redacted for confidentiality appears in square brackets ([]) and is highlighted in red; interview excerpts have also been slightly paraphrased from their original form to preserve anonymity while maintaining their intended meaning.

\begin{figure}[t]
\centering
\begin{center}
\fbox{
\parbox{0.95\linewidth}{
\small
\textbf{User Story:} \\
\textcolor[HTML]{3333FF}{As a \textcolor{red}{[Redacted Role]}, I want to automatically extract and categorize \textcolor{red}{[Internal Messaging Platform]} help requests so that the team leads can identify knowledge gaps and repetitive questions, enabling them to focus more on supporting and developing their team.}
}
}
\end{center}
\caption{User story extracted from the transcript in Fig.~\ref{fig:transcript}, traceable to the interviewee's explicit intent (timestamp 19:26).}
\label{fig:userstory_explicit}
\end{figure}

The transcript in Figure~\ref{fig:transcript} indicates an explicit automation intent: the interviewee requests a centralized view of analyst performance and automated detection of recurring help-request patterns in the internal messaging platform (timestamp 19:26). This explicit intent for automation can be formulated as a user story, as shown in Figure~\ref{fig:userstory_explicit}: The interviewee wants to be able to automatically extract and categorize messaging-platform help requests to identify knowledge gaps and recurring questions for team leads so that they can focus on supporting and developing their team.

\begin{table*}[t]
\caption{Inferred latent requirements from the transcript in Figure~\ref{fig:transcript},  together with the rationale for each requirement.}
\label{tab:automationreq}
\centering
\scalebox{1.1}{
\footnotesize
\begin{tabular}{|p{5cm}|p{10cm}|} 
\hline
\textbf{Latent Requirement} & \textbf{Rationale} \\
\hline
As a \textcolor{red}{[Redacted Role]}, I want to automatically generate structured one-on-one notes on the \textcolor{red}{[Knowledge Sharing Platform]} so that I can ensure consistent documentation quality. & Automating documentation generation helps address manual data entry and formatting of performance metrics, trends, and audit results when documenting one-on-one meeting notes. The activity is currently carried out manually for each analyst by the team lead. The automation can be implemented using \textcolor{red}{[Workflow Management Platform]} through generating structured \textcolor{red}{[Knowledge Sharing Platform]} pages by pre-populating metrics, trends, and audit results from \textcolor{red}{[Data Platform]}. As a result, most of the one-on-one documentation can be automated, ensuring consistent quality across team leads.\\
\hline
As a \textcolor{red}{[Redacted Role]}, I want to automatically transcribe and analyze call recordings for quality indicators so that I can reduce manual call audit time. & Automating call quality screening helps address time-intensive manual listening to call recordings in analyst performance assessment process. The activity is currently carried out manually and takes two hours per analyst monthly. The automation can be implemented using \textcolor{red}{[Internal Chatbot Assistant]} through call transcription and quality analysis. As a result, virtually all of the calls can be screened automatically which will reduce manual audit time by approximately 30-60 minutes per analyst monthly.\\

\hline
\end{tabular}}
\end{table*}

\begin{figure*}[!t]
    \centering    \includegraphics[width=.78\linewidth]{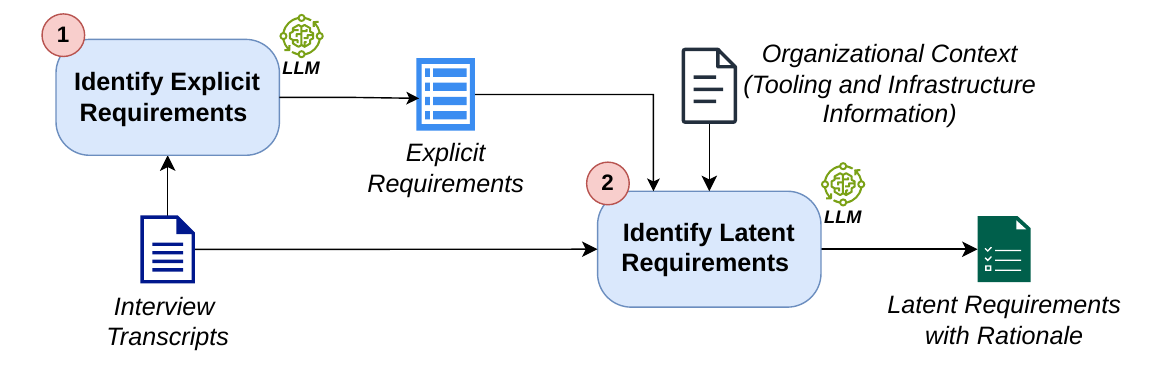}
    \caption{Overview of \framework}
    \label{fig:approach}
\end{figure*}

Moreover, when the pain points identified in the transcript in Figure~\ref{fig:transcript} are considered alongside the capabilities of eSentire's tools and infrastructure, two unstated (latent) requirements can be inferred, as presented in Table~\ref{tab:automationreq}: (i) automatically generating structured one-on-one documentation pre-populated with relevant metrics and trends, and (ii) automatically transcribing and analyzing call recordings for quality indicators. Given the pain point in Figure~\ref{fig:transcript}, the first requirement is suggested by the capabilities of eSentire’s \textcolor{red}{[Knowledge Sharing Platform]} and \textcolor{red}{[Workflow Management Platform]}, while the second is suggested by the capabilities of eSentire’s \textcolor{red}{[Internal Chatbot Assistant]}. Both aim to reduce manual assessment workload. Table~\ref{tab:automationreq} further includes a rationale for each requirement, explaining its importance in eSentire's context and how it could be implemented using existing tools. Each rationale describes the problem addressed, the tools that could implement the requirement, and the \hbox{expected impact of the proposed automation.}

Identifying explicitly stated requirements is relatively straightforward for LLMs, as these are typically conveyed through direct statements of need or intent. Latent requirements, however, are not articulated explicitly and must instead be inferred from descriptions of workflows, pain points, and the broader organizational context. Our goal is to conduct a preliminary but realistic assessment, in an industrial setting, of how well LLMs can perform these two tasks: extracting explicit requirements and inferring plausible latent ones that can also help stimulate further stakeholder discussion in subsequent elicitation iterations by suggesting ideas and opportunities that may not have been initially considered during the interview.

 \section{Our Approach (\framework)}
\label{sec:approach}

Figure~\ref{fig:approach} illustrates \framework, our approach for identifying explicit and latent requirements from stakeholder interview transcripts. \framework\ first extracts explicitly stated requirements from the interview transcripts (Step~1). It then infers additional latent requirements from the transcripts (Step~2), using as inputs the previously identified explicit requirements along with descriptions of the available software tools and automation infrastructure. For each latent requirement, \framework\ also produces a structured rationale that includes the problem addressed, a feasible software solution using the available tools and infrastructure, as well as the expected impact of the resulting automation.

Both steps of \framework\ are LLM-based and rely on prompts with few-shot examples. Table~\ref{tab:promptcomp} presents the elements of the prompts used in these two steps. For example, in the first step, applying \framework\ to the interview excerpt in Figure~\ref{fig:transcript} leads to the generation of the explicit requirement in Figure~\ref{fig:userstory_explicit}. The prompt outline for explicit requirements extraction is provided in Section~\ref{sec:prompt_outline}. Providing \framework\ with a list of tools and infrastructure available at eSentire, together with descriptions of their capabilities, leads to two latent requirements and their rationales in Table~\ref{tab:automationreq} in the second step. The corresponding prompt outline for latent requirements inference is  provided in Section~\ref{sec:prompt_outline}. For both explicit and latent requirements, \framework\ returns the interview timestamp(s) and source text associated with each identified requirement to enable traceability. 
  
\begin{table*}[t]
\centering
\caption{Prompt elements used in each step of \framework.}\label{tab:promptcomp}
\scalebox{1.1}{\begin{tabular}{|p{3cm}|p{9cm}|p{1.2cm}|}
\hline
\textbf{Prompt elements} & \textbf{Brief Description} & \textbf{Used in} \\
\hline
Role Definition & Defines the role of the LLM (extract explicit requirements in Step~1; infer latent requirements and rationale in Step~2). & Steps 1--2 \\
\hline
User Story Criteria & Encodes thirteen quality user story criteria (Table~\ref{tab:qus}) in the literature~\cite{qus} to ensure clarity, completeness, and consistency of generated user stories. & Steps 1--2 \\
\hline
Rules for Explicit Req. & Guides the LLM to extract only explicitly stated requirements and provide traceability to transcript sources. & Step 1 \\
\hline
Rules for Latent Req. & Guides the LLM to infer realistic and plausible  requirements beyond explicit statements. & Step 2 \\
\hline
User Story Template & Provides the standard user story structure (“As a [role], I want [goal], so that [reason]”). & Steps 1--2 \\
\hline
Rationale Template & Specifies a structured justification format including problem, tools, implementation approach, and expected impact ("Automating [activity] helps address [problem/inefficiency] in
[context]. The activity is currently carried out [manually/semi-
automatically/using an existing tool or approach]. The au-
tomation can be implemented using [available tool(s)] through
[implementation approach]. As a result, [expected impact]"). & Step 2 \\
\hline
Few-Shot Examples & Supplies example user stories and rationales to guide consistent and high-quality generation. & Steps 1--2 \\
\hline
Output Format Specification & Defines a structured JSON output format for consistent responses and downstream integration. & Steps 1--2 \\
\hline
Organizational Context (Tooling \& Infrastructure) & Lists available organizational tools and their capabilities to ground feasible requirement generation. & Step 2 \\
\hline
Interview Transcript & Provides stakeholder interview data as the source for extracting explicit and inferring latent requirements. & Steps 1--2 \\
\hline
\end{tabular}}
\end{table*}
 \section{Empirical Evaluation}
\label{sec:empirical_eval}
Our evaluation aims to answer the following questions:

\textbf{RQ1 (Accuracy of Explicit Requirements Extraction)} 
\emph{How accurate is LLM-based extraction of explicit requirements from interview transcripts?}

\textbf{RQ2 (Usefulness of Inferred Latent Requirements)} 
\emph{To what extent do domain experts perceive LLM-derived latent requirements as actionable, clear, and practical?}

\subsection{Interview Transcripts Dataset}\label{sec:transcript}

\begin{table}[t]
\centering
\caption{Summary of the key characteristics of the twelve interview transcripts used in our experiments.}
\label{tab:transcript_summary}

\small

\begin{tabular}{|>{\centering\arraybackslash}m{2.7cm}|c|c|c|c|}
\hline
\textbf{Metric} & \textbf{Mean} & \textbf{Median} & \textbf{Min} & \textbf{Max} \\
\hline
Turns per transcript (\#) & 154 & 145 & 92 & 252\\
\hline
Words per transcript (\#) & 8,938 & 8945 & 6712 & 11987\\
\hline
Transcript duration (minutes) & 59.6 & 59.64 & 44.75 & 79.91\\
\hline
Turn length (words) & 63 & 56 & 41 & 98\\
\hline
\end{tabular}
\end{table}

Our study is based on  a dataset of twelve semi-structured interviews conducted at eSentire as part of their ongoing agile development practices. The interviews covered end-to-end descriptions of real operational processes within a Security Operations Centre (SOC). For example, one interview discussed the analyst performance evaluation process (a sanitized excerpt of which is provided in Figure \ref{fig:transcript}), where multiple stakeholders described how analyst activities are currently tracked, how performance is assessed across different stages, and where bottlenecks occur in consolidating performance-related information. Participants also described specific inefficiencies in the workflow, and discussed desired improvements aimed at increasing automation and reducing operational overhead. All interviews were conducted by eSentire personnel with expertise in requirements elicitation and analysis.

Table~\ref{tab:transcript_summary} summarizes the characteristics of our dataset. Specifically, each interview lasted approximately one hour (mean = 59.6 minutes, range = 45--80 minutes) and included an average of three participants.  Transcripts varied in length, containing between 6,712 and 11,987 words (mean = 8,938) and 92 to 252 speaker turns (mean = 154) suggesting rich, interaction-heavy discussions rather than short survey-style responses. On average, each speaker turn consisted of 63 words, though individual contributions ranged from short clarifications (41 words) to detailed explanations (98 words) with a total speaker turn of 1,237 across all twelve transcripts.

\textcolor{black}{The interview participants represent stakeholders across multiple organizational roles, including analysts, managers, and directors, with varying levels of responsibility. Senior roles account for approximately 36\% of participants, while the remaining 64\% hold junior or intermediate  positions. Together, these roles provide a representative mix of perspectives on day-to-day operations, team coordination, and strategic decision-making within the organization.}

\begin{table*}[t]
\centering
\caption{Requirements evaluation criteria used in \emph{RQ2}, assessing whether the LLM-generated requirements provide time savings compared to manual task execution.}\label{tab:RQ2rating}
\scalebox{1.1}{\begin{tabular}{|p{4.2cm}|p{10cm}|}
\hline
\textbf{Criterion} & \textbf{Description} \\ \hline
\cellcolor{red!40} Negative Time Gain (Reject) & The requirement is not usable; implementing it would take more time than performing the task manually. \\ \hline
\cellcolor{orange!40} No Time Gain (Weak Reject) & The requirement conveys an idea but requires substantial refinement; using it would not result in meaningful time savings. \\ \hline
\cellcolor{yellow!40} Uncertain Time Gain (Borderline) & The requirement is potentially useful, but it is unclear whether it would save time compared to performing the task manually. \\ \hline
\cellcolor{green!30} Moderate Time Gain (Weak Accept) & The requirement is usable with minor refinements; using it would save time overall. \\ \hline
\cellcolor{green!60} High Time Gain (Accept) & The requirement is clear, feasible, and usable; it would save significant time and effort if realized as-is. \\ \hline
\end{tabular}}

\vspace{2.5em}

\caption{Precision, recall, and F1-score for explicit requirements. Highlighted row shows the best F1-score.}
\label{tab:temp_results}
\scalebox{1.1}{\begin{tabular}{|c|c|c|c|c|}
\hline
\textbf{Temperature} & \textbf{Total \# of Req.} & \textbf{Precision (\%)} & \textbf{Recall (\%)} & \textbf{Accuracy (F1) (\%)} \\
\hline
0.0  & 30 & 50 & 94  & 65.3 \\
0.35 & 26 & 64 & 100 & 78.0 \\
\rowcolor{yellow!30}
0.7  & 22 & 73 & 100 & 84.4 \\
1 & 37 & 41 & 94 & 56.5\\
\hline
\end{tabular}}
\end{table*}

\subsection{Experiments and Metrics}
We applied \framework\ to the twelve interview transcripts described in Section~\ref{sec:transcript}. For all experiments, we used Claude Sonnet 4.5 as the underlying LLM. At the time of the study, Claude Sonnet 4.5 was Anthropic's most recent release in the Sonnet family and was recognized for strong reasoning capabilities ~\cite{claude45}. \textcolor{black}{Claude Sonnet 4.5 further supports large context windows, making it well suited for the prompts used in our study~\cite{anthropicContextWindow}}. \textcolor{black}{We considered four different temperature settings, namely 0.0, 0.35, 0.7, and 1.0, following the recommendations provided in the Claude Sonnet 4.5 API documentation~\cite{aimlapi_claude45}, to cover a range of output variability from low (0.0) to highly diverse (1.0). This enabled us to examine how different temperatures within this range affect requirement extraction and latent requirement inference. We limited the study to four temperature levels to keep our evaluation tractable within the scope of this preliminary study.}

To answer \emph{RQ1}, the first author reviewed each transcript to construct a ground-truth (GT) set of explicitly stated requirements. This GT was validated with domain experts to ensure correctness and completeness. The first author then compared the LLM-extracted explicit  requirements from Step~1 of \framework\ with the GT, and computed standard precision-recall metrics as follows: an LLM-generated explicit requirement that matches a requirement in GT is a \emph{true positive}; an LLM-generated explicit requirement that does not match any requirement in GT is a \emph{false positive}; and a GT requirement with no corresponding LLM-generated requirement is a \emph{false negative}. To measure overall accuracy, we use the F1-score, i.e., the harmonic mean of precision and recall.

To answer \emph{RQ2}, we enlisted two domain experts at eSentire (neither of whom is a co-author of this paper) to independently evaluate the LLM-generated latent requirements by completing two tasks. First, the experts evaluated each requirement using a five-point scale to assess whether the automation suggested by the LLM would provide a net benefit in terms of task execution time compared to performing the task entirely manually (Table~\ref{tab:RQ2rating}). 
\textcolor{black}{We mapped the five-point rating scale used in prior work~\cite{voria2025recover} (Reject, Weak Reject, Borderline, Weak Accept, and Accept) to an interpretation based on estimated time savings, since our assessment focuses on time gain rather than general acceptance. The scale ranges from Negative Time Gain (Reject), where the automation would take longer than performing the task manually (including any time needed to correct automation errors) to High Time Gain (Accept), where the automation is perceived to significantly reduce task completion time.}

Second, the experts evaluated the latent requirements along two dimensions: (a) \emph{novelty} (whether they provided previously unconsidered insights) and (b) \emph{feasibility} (whether they were realistic and technically implementable with the organization's available tools and infrastructure). The experts could also provide requirement-specific concerns or refinement suggestions via an open-ended comment field.

Both experts had substantial practical experience with eSentire's organizational structure, operational processes, and tool ecosystem. Before the evaluation, we met with each expert individually to review the context, objectives, procedure, and evaluation criteria. They then completed the evaluation independently at their own pace, spending about four to five hours each reviewing the materials and providing feedback.

\subsection{RQ1 (Accuracy)}\label{sec:rq1}

Table~\ref{tab:temp_results} presents the precision, recall, and F1-score from applying \framework\ to the twelve transcripts at temperatures 0.0, 0.35, 0.7, and 1.0. At 0.0, \framework\ extracted 30 requirements with 50\% precision, while at 1.0 it extracted 37 requirements with 41\% precision and 94\% recall for both. Intermediate temperatures (0.35 and 0.7) improved precision (64\% and 73\%, respectively) while yielding perfect recall, with temperature 0.7 achieving the highest F1-score (84.4\%) among the evaluated settings. These results suggest that both very low and very high temperature settings can degrade extraction quality, while intermediate values provide better overall performance within the range of temperatures explored in this study.

\begin{table*}[t]
\centering
\caption{Percentage distribution of expert ratings for LLM-identified latent requirements based on estimated time gains.}
\label{tab:acceptability_percentage}
\scalebox{1.1}{\begin{tabular}{|c|p{2.4cm}|p{2cm}|p{2.6cm}|p{2.6cm}|p{2cm}|}
\hline
\textbf{Rater} & \cellcolor{red!40}\textbf{Negative Time Gain} & \cellcolor{orange!40}\textbf{No Time Gain} & \cellcolor{yellow!40}\textbf{Uncertain Time Gain} & \cellcolor{green!30}\textbf{Moderate Time Gain} & \cellcolor{green!60}\textbf{High Time Gain} \\
\hline
Expert 1 & 8.7\% & 3.8\% & 14.4\% & 46.2\% & 26.9\% \\
\hline
Expert 2 & 15.4\% & 1.0\% & 6.7\% & 20.2\% & 56.7\% \\
\hline
\end{tabular}}
\end{table*}

We manually inspected the false positives and grouped them into three types based on the linguistic cues that may have led \framework\ to infer them incorrectly. Figure~\ref{fig:FP} summarizes these categories, their frequencies, and an example false positive for each category with the excerpt from which it was extracted. First, mentions of manual work -- without any stated intent to automate -- were sometimes misclassified as explicit requirements (66\% of observed false positives). This occurred despite the  rules in our prompts (Table~\ref{tab:promptcomp}), which instruct the LLM not to infer requirements from manual work alone when no explicit automation intent is mentioned. Note that the manual work example in Figure~\ref{fig:FP} could imply a useful latent requirement, but since it is not explicit, it constitutes a false positive for RQ1. Second, references to existing automation sometimes led the LLM to treat them as future needs and requirements (17\%). Third, confirmation-seeking conversational cues (e.g., ``right?'') about automation that already exists could be misinterpreted as uncertainty, causing the LLM to \hbox{find a requirement where none existed (17\%).}

\begin{figure}[t]
\centering
\setlength{\fboxsep}{6pt}
\small
\scalebox{1}{\fbox{\begin{minipage}{0.95\columnwidth}

\textcolor{blue}{\textbf{1. Mentions of Manual Work Mistaken for Requirements (66\% of FPs)}}\\
\textbf{Example:} \emph{Interview Excerpt:} Reviewing analyst performance data is currently done manually to identify key issues, which can be time-consuming.

\emph{Explicit Requirement:} As a \textcolor{red}{[Redacted Role]}, I want to automate the review and synthesis of analyst performance data, so that I can efficiently provide meaningful feedback without spending significant time.

\vspace{6pt}\hrule\vspace{6pt}

\textcolor{blue}{\textbf{2. Discussion of Existing Automation Mistaken for Requirements (17\% of FPs)}}\\
\textbf{Example:} \emph{Interview Excerpt:} Our \textcolor{red}{[Internal Process]} is done by AI and it can highlight areas for improvement mid-cycle.

\emph{Explicit Requirement:} As a \textcolor{red}{[Redacted Role]}, I want AI to highlight mid-cycle \textcolor{red}{[Internal process]} insights, so that I can identify areas for improvement before the cycle ends.

\vspace{6pt}\hrule\vspace{6pt}

\textcolor{blue}{\textbf{3. Confirmation-Seeking Language Mistaken for Requirements (17\% of FPs)}}\\
\textbf{Example:} \emph{Interview Excerpt:} The \textcolor{red}{[Internal Process]} that was previously handled manually is now automated, right?

\emph{Explicit Requirement:} As a \textcolor{red}{[Redacted Role]}, I want the \textcolor{red}{[Internal process]} to be automated, so that handling \textcolor{red}{[Internal process]} is more efficient and consistent.

\end{minipage}}}
\caption{Categories of false positives (FPs) in explicit requirements, with proportions and examples.}
\label{fig:FP}
\end{figure}

\textit{The answer to RQ1 is that} \framework\ can extract explicitly stated  requirements with high accuracy -- achieving its best results at a temperature of 0.7 (precision 73\%, recall 100\%, F1 84.4\%). False positives were attributable to the LLM misinterpreting mentions of manual work, discussions of existing automation, or confirmation-seeking language as  requirements.

\subsection{RQ2 (Usefulness)} \label{sec:rq2}
For RQ2, domain experts evaluate the usefulness of 104 latent requirements generated by \framework\ at a temperature of 0.7, the best-performing among the evaluated settings in RQ1. Table~\ref{tab:acceptability_percentage} shows the percentage distribution of expert ratings based on estimated time gains compared to manual task execution. As shown in the table, an average of 75\% of the requirements were rated as providing positive time gains (Moderate, or High Time Gain), indicating that most of the LLM suggested requirements are likely to save time. 14.5\% were rated as having negative or no time gain, suggesting that relatively few requirements were perceived as impractical or not worth automating. 10.5\% were rated as uncertain in terms of time benefit.

We use Cohen's kappa ($\kappa$)~\cite{cohen1960kappa} for measuring inter-rater agreement. Collapsing High/Moderate Time Gain and Negative/No Time Gain into binary categories yields a $\kappa$ of 0.54, indicating moderate agreement. Agreement occurs when both experts assign the same category to a requirement, while disagreement occurs when they assign different categories.

\textcolor{black}{It is important to note that the percentage of requirements associated with positive time gains (75\%) does not imply that these requirements are ready to be added directly to a production backlog. Rather, the evaluation measures whether experts viewed the generated requirements as useful starting points for identifying automation or process-improvement opportunities that could save time if further explored. In practice, inferred latent requirements would still require stakeholder discussion, negotiation, feasibility assessment, and refinement before being considered implementation-ready.}

\begin{figure*}[t]
  \centering
  \noindent
  \scalebox{1}{\fbox{\begin{minipage}{1\linewidth}
      \textcolor{blue}{\textbf{1. Lack of concrete technical details in the rationale}} {\color{orange}($\approx$\,49\% of all feedback)} \\
      \textbf{Example feedback:} 
      \textit{``No implementation details,” “Not enough details,” “Unclear where data is stored.''}\\
      \textbf{Insight:} Although requirements are conceptually sound, the rationale often provides insufficient reasoning to justify feasibility.

      \medskip
      \hrule
      \medskip

      \textcolor{blue}{\textbf{2. Practical applicability depends on access to specific data sources or APIs.}}   {\color{orange}($\approx$\,22\% of all feedback)}\\  
      \textbf{Example feedback:}
      \textit{``Possible if it has access to…,” “Can work granted API access…''}\\
      \textbf{Insight:} Actionability depends strongly on available data and system connectivity.

      \medskip
      \hrule
      \medskip

      \textcolor{blue}{\textbf{3. Proposed tool functionality exceeds the realistic capabilities.}} {\color{orange}($\approx$\,16\% of all feedback)}\\
      \textbf{Example feedback:}
      \textit{``Not Possible with \textcolor{red}{[Workflow Management Platform]}…,” “\textcolor{red}{[Internal Chatbot Assistant]} can't do this…''}\\
      \textbf{Insight:} Misalignment between generated requirements and platform functionality.

      \medskip
      \hrule
      \medskip

      \textcolor{blue}{\textbf{4. Tasks can be performed with conventional tools without AI.}} {\color{orange}($\approx$ 7\% of all feedback)}\\
      \textbf{Example feedback:}
      \textit{``Better ways to do this without AI,” “This is a job for a security tool.''}\\
      \textbf{Insight:} \framework\ sometimes over-relies on AI as a default solution, even when simple approaches would be sufficient.
    \end{minipage}}}
  \caption{Key themes in expert feedback on latent requirements.}
  \label{fig:expert-feedback-themes}
\end{figure*}

Figure~\ref{fig:expert-feedback-themes} summarizes the main themes in expert feedback on latent requirements and their rationale. As shown in the figure, the most frequent feedback concerns the lack of concrete technical details in the rationale ($\approx$49\%), where experts noted that requirements were conceptually reasonable but lacked sufficient implementation detail to justify feasibility. The second most common theme relates to dependencies on specific data sources or APIs ($\approx$22\%), indicating that the practicality of several requirements depends on access to particular systems or integrations. A smaller portion of feedback highlights misalignment with the capabilities of the proposed tools ($\approx$16\%), where the suggested automation exceeds what the referenced platforms can realistically support. A minority of comments ($\approx$7\%) note that some tasks could be addressed using conventional tools rather than AI, suggesting occasional over-reliance on AI-based solutions. Finally, a small number of isolated remarks ($\approx$6\%) did not fit any of these themes.

Regarding novelty and feasibility, both experts observed that the LLM-derived latent requirements were often novel and capable of suggesting requirements that stakeholders may not have initially considered. \textcolor{black}{At the same time, expert feedback suggests that the usefulness of the generated latent requirements is highly context-dependent, especially in terms of implementation feasibility and expected impact; nevertheless, experts still perceived them as practically valuable.
}

\textit{The answer to RQ2 is that} \framework\ generates latent  requirements that experts generally perceive as useful and novel (75\% rated Moderate/High time gain), but feasibility is sometimes unclear due to missing context (e.g., data/API access, tool details), limited technical rationale, occasional tool-capability mismatches, and overuse of AI.

\subsection{Threats to Validity}

\textbf{Internal validity.} \textcolor{black}{Our evaluation involved human judgment at several stages. For RQ1, the first author constructed the ground-truth set of explicitly stated requirements and matched the requirements extracted by LENS against this set. This process may introduce researcher bias or inconsistency, particularly in borderline cases where a stakeholder statement could be interpreted either as an explicit requirement or as a description of an existing process or problem. We mitigated this threat by adopting a narrow definition of explicit requirements and having the ground truth reviewed and validated by domain experts. However, some subjectivity may remain and could affect the reported results. For RQ2, the assessments usefulness, novelty, feasibility, and expected time gain may have reflected the experts' individual experience and implementation assumptions. We reduced this threat by involving two domain experts who were not co-authors, clarifying the evaluation criteria in advance, and collecting their assessments independently. The moderate inter-rater agreement nevertheless indicates residual subjectivity. The RQ2 results should therefore be interpreted as measures of perceived usefulness and feasibility, rather than conclusive evidence of realized time savings or implementation success.}

\textbf{External validity.} \textcolor{black}{The interview transcripts were collected from a single organization, and the evaluation of latent requirements relied on judgments from only two domain experts within that organization. However, the experts have substantial domain knowledge, and the dataset spans diverse roles, topics, and operational challenges, making it suitable for a preliminary evaluation. Nonetheless, differences in application domains, organizational practices, interview styles, and the number and composition of expert evaluators may limit the generalizability of our findings. We used Claude Sonnet 4.5~\cite{claude45} because it matched eSentire's deployment environment at the time and is a production-grade LLM. Results may differ for other LLMs due to variations in their capabilities and behaviour.} \section{Related Work}
\label{sec:relwork}

Recent work has explored the use of LLMs to process stakeholder interview transcripts for requirements engineering. RECOVER~\cite{voria2025recover} leverages natural language processing and LLMs to identify requirements-relevant dialogue and generate system requirements from explicit stakeholder statements. In their study, out of 133 speaker turns, 62 were predicted as requirement-relevant while 71 were classified as not relevant. Spijkman et al.~\cite{reconsum2023} propose REConSum, which summarizes stakeholder conversations by filtering speaker turns and retaining only those that include a question and a subsequent answer likely to contain requirements-relevant information. While both approaches share our overarching goal of leveraging LLMs to automatically extract requirements from interview transcripts, their primary emphasis is on identifying and consolidating requirement-relevant speaker turns. In contrast, our approach does not focus on predicting requirement-relevant turns. Instead, we use full transcripts as a foundation for identifying both explicit and latent requirements.

Jain et al.~\cite{jain2023transformer} address the challenge of interpreting software engineering contracts, which are often lengthy and written in complex legal terminology that is difficult to understand and interpret correctly. They propose an approach that summarizes such contracts into a form more comprehensible to requirements analysts. While this work focuses on contracts and their summarization, our work, although sharing the broader goal of extracting requirements, is based on the hypothesis that LLMs can be leveraged beyond summarization tasks to support inference of latent requirements not explicitly stated by the stakeholders.

Shrafudin et al. \cite{sharfuddin2025generative} take a complementary yet distinct direction by extracting stakeholder goals from interview transcripts and representing them as structured goal models using textual entailment and LLM-based inference. Their work aims to support analysis through goal modeling. Unlike this goal-centric perspective, our approach extends beyond summarization and represent both explicit and latent requirements as user stories, along with supporting rationale.

\textcolor{black}{Our study is conceptually related to prior work on creativity in requirements engineering, which seeks to generate novel and useful requirements ideas. Maiden et al.~\cite{maiden2004provoking} define creative requirements as those that are both novel and appropriate, and report how structured creativity workshops can support the generation of new requirements and design ideas. Rather than facilitating creativity within live stakeholder workshops, our work examines whether an LLM can infer latent requirements from completed stakeholder interview transcripts and organizational context.}

 \section{Lessons Learned}
\label{sec:lessons}

\vspace*{.1em}\textbf{\textit{Lesson 1: Conversational cues can cause LLMs to misinterpret intent:}} 
Although \framework\ performed well in extracting explicit requirements (73\% precision and 100\% recall), our findings show that LLMs can misclassify requirements when stakeholder language contains confirmation-seeking cues, such as “right?”. In such cases, the LLM inferred automation intent for tasks that were already automated, leading to false positives (Section~\ref{sec:rq1}). Similar observations have been reported in prior work~\cite{voria2025recover}, where misclassifications frequently arise from non-informative conversational utterances.

\textbf{\textit{Implication:}} A light pre-processing step that removes or neutralizes confirmation-seeking language (e.g., “right?”) may reduce misinterpretations caused by conversational noise.

\vspace*{.1em}\textbf{\textit{Lesson 2: LLMs may misassess technical feasibility:}} 
Although \framework\ generated latent requirements that experts rated as providing positive time gains in approximately 75\% of cases, expert feedback in Section~\ref{sec:rq2} indicates that feasibility may remain unclear when the generated rationales lack technical detail, assume unavailable integrations or data access, or exceed the capabilities of existing tools.
\textcolor{black}{This points to the importance of organizational context in shaping the quality and feasibility of latent requirements. Without sufficient grounding in the organization's tools, processes, and data sources, LLM-generated suggestions may appear plausible while remaining misaligned with actual system constraints.}

\textbf{\textit{Implication:}} Improving the feasibility of automatically inferred latent requirements requires rich descriptions of the organization's tools, infrastructure, and available data sources, ideally developed in collaboration with domain experts.

\textbf{\textit{Lesson 3: Expert review remains key:}} 
Section~\ref{sec:rq2} showed only moderate agreement among experts, suggesting that the usefulness of inferred requirements may depend on contextual factors that are inherently tacit or are too difficult or costly to fully articulate. \textcolor{black}{Inferred latent requirements should therefore not be interpreted as finalized or implementation-ready, instead, they are better viewed as ideas that can guide subsequent rounds of stakeholder discussion.} 

\textcolor{black}{\textbf{\textit{Implication:}} LLM-derived latent requirements can be valuable starting points for ideas or opportunities that stakeholders may not have initially considered. However, these requirements should be carefully vetted by domain experts and treated as inputs to an iterative requirements elicitation process rather than as final and validated outcomes.}

\section{Conclusion}\label{sec:conclusion}

We proposed \framework, an approach that extracts explicitly stated requirements from stakeholder interview transcripts and infers latent ones by interpreting stakeholder statements in the context of organizational workflows, tools, and operational constraints. Our preliminary evaluation with eSentire shows that LLMs can provide practical support for both tasks: explicit requirements were extracted with an F1-score of 84.4\%, and 75\% of the inferred latent requirements were judged by domain experts to offer time-saving potential. With respect to inferred latent requirements, our results further suggest that their main value is not in producing ready-to-implement features, but rather in discovering opportunities to reframe stakeholder conversations around mitigating inefficiencies and realizing untapped automation potential.

Given the preliminary nature of our work, the findings should be interpreted only as early evidence rather than general claims. Our study is based on one organization, one domain, and a limited set of interviews. In future work, we would like to evaluate our approach with more organizations and operational settings. We would also like to further qualify our definition of latent requirements. While the current definition captures requirements that are plausible in light of workflows, tools, and infrastructure, it needs to be strengthened to also consider whether such requirements are ethically appropriate, aligned with organizational guardrails, compliant with laws and regulations, and acceptable to stakeholders.

\section*{Acknowledgements}
Financial support for this research was provided by Mitacs, eSentire, and NSERC of Canada under the Discovery program. We are grateful to the staff at eSentire, particularly Taha Ansari, for his guidance and support throughout the research.

\section*{Data Availability }
\textcolor{black}{The underlying dataset and complete artifacts cannot be publicly released due to confidentiality constraints. To support methodological replication, the paper provides the workflow, prompt outlines, templates, LLM settings, evaluation protocol, and aggregate results.} \section{appendix} \label{sec:appendix}
\textcolor{black}{This appendix presents the prompt outlines for both steps of \framework\ \textcolor{black}{(full prompts omitted due to confidentiality constraints)}, along with the thirteen QUS criteria~\cite{qus} used to guide the LLM in generating well-structured and consistent user stories.}
\subsection{Prompt Outline}\label{sec:prompt_outline}

\noindent
\fbox{
\begin{minipage}{0.97\linewidth}
\textbf{Prompt Outline for Explicit Requirements Extraction}

\vspace{0.2em}
\textbf{(I) Role:} Explicit requirements extractor

\vspace{0.2em}
\textbf{(II) Input:} Interview transcript\newline
13 QUS Criteria from Table \ref{tab:qus} \newline
Rules for explicit requirements extraction\newline
\textcolor{blue}{\{User story template\}} presented in Table \ref{tab:promptcomp}\newline
Few-shot examples\newline
Output format specification

\vspace{0.2em}
\textbf{(III) Output:} Explicit requirements represented as user stories along with timestamp and corresponding excerpt from transcript

\end{minipage}
}

\vspace{1em}
\noindent
\fbox{
\begin{minipage}{0.97\linewidth}
\textbf{Prompt Outline for Latent Requirements Inference}

\vspace{0.2em}
\textbf{(I) Role:} Latent requirements identifier

\vspace{0.2em}
\textbf{(II) Input:} Interview transcript\newline
Explicit requirements extracted from the previous step\newline
13 QUS Criteria from Table \ref{tab:qus} \newline
Rules for identifying latent requirements \newline
\textcolor{blue}{\{User story template\}} presented in Table \ref{tab:promptcomp} \newline
\textcolor{blue}{\{Rationale template\}} presented in Table \ref{tab:promptcomp} \newline
Few-shot examples\newline
Organizational context (tooling \& infrastructure)\newline
Output format specification

\vspace{0.2em}
\textbf{(III) Output:} Latent requirements represented as user stories with timestamp and corresponding excerpt from transcript

\end{minipage}
}

\subsection{QUS Criteria}

\begin{table}[H]
\centering
\caption{Quality user story (QUS) criteria~\cite{qus} used in the prompts (see Table~\ref{tab:promptcomp}) } \label{tab:qus} 
\scalebox{1.15}{\begin{tabular}{|p{1.1cm}|p{5.8cm}|}
\hline
\textbf{Category} & \textbf{Criteria} \\
\hline
\textbf{Syntactic} & 
\textbf{Atomic}: Each story expresses exactly ONE feature.\\
& \textbf{Minimal}: Contains only role, means, and ends.\\
& \textbf{Well-formed}: Includes at least a role and a means.\\
\hline
\textbf{Semantic} & 
\textbf{Conflict-free}: Not inconsistent with other user stories.\\
& \textbf{Conceptually sound}: The means expresses a feature, the ends expresses a rationale.\\
& \textbf{Problem-oriented}: Specifies the problem, NOT the solution.\\
& \textbf{Unambiguous}: Avoids terms that may lead to multiple interpretations.\\
\hline
\textbf{Pragmatic} & 
\textbf{Full sentence}: Each story is represented as a well-formed full sentence \\
& \textbf{Complete}: Together, all stories create a feature-complete solution.\\
& \textbf{Independent}: Self-contained, avoiding inherent dependencies.\\
& \textbf{Estimatable}: Not too coarse-grained, easy to plan and prioritize.\\
& \textbf{Uniform}: All stories follow roughly the same template.\\
& \textbf{Unique}: No duplicates.\\
\hline
\end{tabular}}
\end{table}

\balance
\bibliographystyle{plain}
\bibliography{references}
\end{document}